# Tilted spin current generated by the collinear antiferromagnet RuO$_2$

Arnab Bose*[1,2], Nathaniel J. Schreiber*[3], Rakshit Jain*[1,2], Ding-Fu Shao[4], Hari P. Nair[3], Jiaxin Sun[3], Xiyue S. Zhang[1], David A. Muller[1,5], Evgeny Y. Tsymbal[4], Darrell G. Schlom[3,5,6], and Daniel C. Ralph[2,5†]

1 School of Applied and Engineering Physics, Cornell University. Ithaca, NY 14853, USA

2 Department of Physics, Cornell University. Ithaca, NY 14853, USA

3 Department of Materials Science and Engineering, Cornell University. Ithaca, NY 14853, USA

4 Department of Physics and Astronomy & Nebraska Center for Materials and Nanoscience, University of Nebraska, Lincoln, Nebraska 68588-0299, USA

5 Kavli Institute at Cornell for NanoScale Science, Ithaca, NY 14853, USA

6 Leibniz-Institut für Kristallzüchtung, Max-Born-Straße 2, 12489 Berlin, Germany

* These authors contributed equally

† dcr14@cornell.edu

**Abstract**

We report measurements demonstrating that when the Néel vector of the collinear antiferromagnet RuO$_2$ is appropriately canted relative to the sample plane, the antiferromagnet generates a substantial out-of-plane damping-like torque. The measurements are in good accord with predictions that when an electric field ($\vec{E}$) is applied to the spin-split band structure of RuO$_2$ it can cause a strong transverse spin current even in the absence of spin-orbit coupling. This produces characteristic changes in all three components of the $\vec{E}$-induced torque vector as a function of the angle of $\vec{E}$ relative to the crystal axes, corresponding to a spin current with a well-defined tilted spin orientation $\hat{s}$ approximately (but not exactly) parallel to the Néel vector, flowing perpendicular to both $\vec{E}$ and $\hat{s}$. This angular dependence is the signature of an antiferromagnetic spin Hall effect with symmetries that are distinct from other mechanisms of spin-current generation reported in antiferromagnetic or ferromagnetic materials.

**Introduction**

Symmetries play a central role in determining the form of electric-field-induced spin currents, and how those spin currents generate spin-transfer torques in magnetic devices. In most materials, rotational and reflection symmetries require that electric-field-induced spin currents have spins allowed only along a single axis, in-plane and perpendicular to $\vec{E}$. This is unfortunate, because this orientation of spin cannot drive efficient anti-damping switching of magnetic devices with perpendicular magnetic anisotropy. If the symmetry of the spin-current-generating material is lowered, more-general spin orientations are possible. This has been demonstrated in materials for which the symmetry is lowered by the crystal structure[1–4] or a ferromagnetic moment[5–14], although the resulting out-of-plane anti-damping torques have not yet been demonstrated to be strong enough for practical applications. Spin currents with out-of-plane spin components emerging from antiferromagnets have also been predicted and measured[15–22]. This is exciting, because the spin-current generation in antiferromagnets appears to result primarily from different mechanisms – exchange interactions and Berry curvature induced by non-collinear spin order – that have the potential to be stronger than the spin-orbit interactions that had been the previous focus of this field.

Recently, González-Hernández *et al.*[23] proposed a previously-unrecognized mechanism whereby the collinear antiferromagnet $RuO_2$ might produce strong electric-field-induced spin currents with spin orientation approximately aligned along the Néel vector. In this model, a spin-split band-structure resulting from inequivalent crystal environments for the two spin sub-lattices naturally provides spin-splitter functionality to yield spin currents transverse to an applied electric field. When the Néel vector is appropriately canted, the result can be a vertically-flowing spin current with a strong out-of-plane spin component. Here, we test this prediction by measuring $\vec{E}$-induced torques within $RuO_2$/Permalloy (Py) devices. We demonstrate a method by which the full 3-dimensional spin-orientation of the $\vec{E}$-induced spin can be determined by measuring the $\vec{E}$-induced torques as a function of the angles of both the applied magnetic field and $\vec{E}$ relative to the crystal axes. We find results in good agreement with the spin-splitter mechanism for $RuO_2$, with an $\vec{E}$-induced spin-current polarization aligned approximately (but not exactly) with the Néel vector.

It is important to note that the mechanism we measure is different from a recent report of a spin Hall effect (SHE) in the collinear antiferromagnet $Mn_2Au$[21]. In $Mn_2Au$, the spin polarization of the spin current is perpendicular to the Néel vector $\mathcal{N}$, and the out-of-plane anti-damping torque is maximized when the $\vec{E}$ is parallel to $\mathcal{N}$. In contrast, for the spin-splitter mechanism and for our measurements on $RuO_2$, the spin polarization of the spin current is approximately parallel to $\mathcal{N}$, and the existence of an out-of-plane spin current requires that $\mathcal{N}$ to be canted out of the sample plane. Our angle-dependence analysis also achieves fully quantitative measurements of the resulting unconventional spin-transfer torques, something that was not attempted in the previous study.

**Background**

$RuO_2$ is a conductive rutile oxide belonging to the space group $P\frac{4_2}{m}nm$ (no. 136), with a globally centrosymmetric crystal structure (Fig. 1(a)). The crystal structure is the same as $IrO_2$, the spin Hall effect of which has been measured recently[24], although the two materials differ in that $RuO_2$ is an antiferromagnet while $IrO_2$ is non-magnetic. The band structures of both materials contain Dirac nodal lines protected by nonsymmorphic symmetries[25,26], which have the potential to generate strong SHEs. Recently $RuO_2$ has also drawn a considerable attention with the discovery of strain stabilized superconductivity[27,28], anomalous antiferromagnetism[29,30], a crystal Hall effect[31], and the predictions of a magnetic spin Hall effect[23] and a giant tunneling magnetoresistance effect in a $RuO_2$-based antiferromagnetic tunnel junctions[32,33].

Within the structure of $RuO_2$ (Fig. 1(a), the Ru atoms at the center and corners of the unit cell have opposite spins, forming a collinear antiferromagnetic (AF) ordering with the Néel vector ($\mathcal{N}$) oriented along the [001] or [00$\bar{1}$] axis (with perhaps a small canting)[29,30]. The center and the corner Ru atoms experience different oxygen environments, which leads to spin-split bands as shown schematically in Fig. 1(b)[23]. Due to this spin-split band structure, theory predicts that when an electric field ($\vec{E}$) is applied it can generate a transverse spin current through an AF spin Hall effect (AF-SHE), even in the absence of any spin-orbit interaction (SOI). (Fig. 1(c) shows the Fermi surface shifted by an applied field, compared to the $\vec{E} = 0$ Fermi surface in Fig. 1(b).) In general, the flow of the spin current from this AF-SHE will have a spin orientation $\hat{s}$ pointing approximately along $\mathcal{N}$ with a spin-flow direction perpendicular to both $\vec{E}$ and $\hat{s}$. The result can be qualitatively similar to the conventional SHE in heavy metals, which is even under time reversal ($\mathcal{T}$-even), but the origin of the AF-SHE is fundamentally different because it does not require SOI and it is $\mathcal{T}$-odd. Our density functional theory calculation results shown in Fig. 1(d) predict that in a single-domain RuO$_2$ antiferromagnet the $\mathcal{T}$-odd spin Hall conductivity (SHC) should be almost an order of magnitude larger than the $\mathcal{T}$-even SHC, given the SOI of RuO$_2$ (see the Supplemental Materials for details). This result is consistent with the prediction of González-Hernández *et al*.[23] A $\mathcal{T}$-odd transverse spin current in magnetic systems is generally referred to as a magnetic SHE[15,18,34].

Based on the geometry of this AF-SHE, we expect very different symmetries for the $\vec{E}$-induced torques depending on whether the Néel vector is oriented out-of-plane or canted partially into the sample plane. For (001)-oriented RuO$_2$ films, $\mathcal{N}$ and hence the $\vec{E}$-generated spin-orientation $\hat{s}$ should point approximately perpendicular to the sample plane (Fig. 1(e)). Since the flow of spin current generated by the AF-SHE must be perpendicular to $\hat{s}$, the AF-SHE mechanism should not generate any spin current flowing vertically toward the Py layer, and hence should not produce any spin-transfer torque in the (001)-oriented case. Therefore, for this geometry there should be only a conventional $\mathcal{T}$-even SHE induced by SOI, with spin oriented in-plane and perpendicular to $\vec{E}$, *i.e.*, in the Y-direction as defined in Fig. 1(e). This can lead to both an in-plane damping-like torque with the symmetry $\vec{\tau}_{DL}^Y = -\tau_{DL,0}^Y(\hat{m} \times \hat{Y} \times \hat{m})$ and/or an out-of-plane field-like torque $\vec{\tau}_{FL}^Y = -\tau_{FL,0}^Y(\hat{m} \times \hat{Y})$. (We define signs as in Ref. [35]; with these choices of sign, a positive $\tau_{DL,0}^Y$ corresponds to the sign of the spin Hall effect for Pt, and $\tau_{FL,0}^Y$ is positive for the Oersted field. All torques quoted are per unit magnetic moment with sign $\vec{\tau} \propto d\hat{m}/dt$.)

For a (101)-oriented RuO$_2$ film, on the other hand, $\mathcal{N}$ is canted away from the out-of-plane direction with components along both the out-of-plane [101]* and in-plane [$\bar{1}$01] directions (Fig. 1(e,f)). The AF-SHE in this geometry should generate a vertically-flowing spin current with $\hat{s}$ having both out-of-plane and in-plane components. If we define the X direction as parallel to the applied electric field $\vec{E}$ and the Z direction as out-of-plane, then depending on the orientation of $\vec{E}$ relative to the crystal axes the AF-SHE can produce a spin Hall conductivity with non-zero components $\sigma_{ZX}^Y$ (generating, e.g., in-plane damping-like torque), $\sigma_{ZX}^X$ (generating, e.g., $\vec{\tau}_{DL}^X = \tau_{DL,0}^X(\hat{m} \times \hat{X} \times \hat{m})$) and $\sigma_{ZX}^Z$ (generating, e.g., $\vec{\tau}_{DL}^Z = \tau_{DL,0}^Z(\hat{m} \times \hat{Z} \times \hat{m})$). A symmetry analysis of the SHC tensor (see Supplemental Material) predicts that the components of damping-like torque generated by the AF-SHE will follow a signature angular-dependence pattern $\tau_{DL,0}^\alpha \propto (\hat{E} \times \hat{s}) \cdot \hat{Z}\, s_\alpha$. In a (101)-oriented RuO$_2$ film if the $\vec{E}$-generated spin $\hat{s}$ is oriented at an angle $\theta_s$ tilted away from the out-of-plane (Z = [101]*) axis toward the [$\bar{1}$01] direction and $\hat{E}$ is applied at an angle $\psi$ with respect to the in-plane [010] axis, this means that the components of damping-like torque efficiencies per unit electric field generated by the AF-SHE should follow

$$\xi_{DL,E}^Y \equiv \frac{2e}{\hbar} \frac{\mu_0 M_s t_{FM}}{\gamma} \frac{\tau_{DL,0}^Y}{E} = C_1 \sin^2\theta_s \cos^2\psi + C_0 \quad (1)$$

$$\xi_{DL,E}^Z \equiv \frac{2e}{\hbar} \frac{\mu_0 M_s t_{FM}}{\gamma} \frac{\tau_{DL,0}^Z}{E} = C_1 \cos\theta_s \sin\theta_s \cos\psi \quad (2)$$

$$\xi_{DL,E}^X \equiv \frac{2e}{\hbar} \frac{\mu_0 M_s t_{FM}}{\gamma} \frac{\tau_{DL,0}^X}{E} = C_1 \sin^2\theta_s \cos\psi \sin\psi, \quad (3)$$

where $C_0$ and $C_1$ are constants, $M_S$ in the saturation magnetization, $t_{FM}$ is the thickness of the ferromagnetic layer, and ɣ is the gyromagnetic ratio. (The $C_0$ term can be generated by the conventional SHE, independent of the AF-SHE.)

In micrometer-scale devices of the sort we study, we expect to have many AF domains, with nearly equal numbers having $\mathcal{N}$ pointing along [001] and [00$\bar{1}$]. If the fractions of the two domain variants are exactly equal, this should lead to zero $\mathcal{T}$-odd spin current. Nonetheless, in a multi-domain scenario, we can still obtain either a non-zero $\mathcal{T}$-odd SHC due to imperfect cancellation between domains or a non-zero $\mathcal{T}$-even SHC due to non-zero SOI. We will refer to a non-trivial spin Hall torque with a well-defined spin orientation $\hat{s}$ (and therefore following the signature angular dependence in Eqs. (1)-(3)) as an AF-SHE for either the $\mathcal{T}$-odd or $\mathcal{T}$-even mechanism.

**Experimental techniques**

To investigate the spin Hall conductivity of RuO$_2$, we measure the $\vec{E}$-induced torques acting on an in-plane-magnetized Ni$_{80}$Fe$_{20}$ (Py) layer (4 nm) deposited by sputtering on top of epitaxial RuO$_2$ thin films (3-9 nm). We analyze films grown on both (001)- and (101)-oriented TiO$_2$ substrates in order to study two different orientations of the Néel vector relative to the sample plane. Details of the sample preparation are provided in the Supplemental Material.

We measure the $\vec{E}$-generated torques in Py/RuO$_2$ bilayers using angle-dependent spin-torque ferromagnetic resonance (ST-FMR)[1,36,37], and we verify the results using the in-plane-anisotropy harmonic Hall (HH)[38] technique. For ST-FMR measurements, the Py/RuO$_2$ bilayer is patterned into a wire 30 μm × 20 μm with microwave-compatible contacts as shown in Fig. 1(g). On each sample chip we make several such devices oriented at different angles relative to the crystal axes to vary the angle $\psi$ of $\vec{E}$ relative to the [010] direction (see Methods for more details). We apply a fixed-frequency microwave current and sweep an in-plane magnetic field at an angle $\phi$ with respect to the current flow direction (Fig. 1(g)), repeating the measurement for each sample for field angles from $\phi = -90°$ to 270° in steps of 15°. Torques acting on the magnet from the microwave current induce magnetic precession and corresponding resistance oscillations due to the bilayer's anisotropic magnetoresistance. Mixing between the resistance oscillations and the oscillating current results in DC voltage signals near resonance, the magnetic-field dependence of which can be fit to the sum of symmetric and antisymmetric Lorentzian peak shapes[1,36,37]: $V_S = S\left(\frac{\Delta^2}{(H-H_0)^2+\Delta^2}\right)$ and $V_A = A\left(\frac{(H-H_0)\Delta}{(H-H_0)^2+\Delta^2}\right)$ (Fig. 2(a)). The amplitudes $S$ and $A$ are related to the in-plane and out-of-plane $\vec{E}$-induced torques, $\Delta$ is the linewidth, and $H_0$ is the resonant field. In the most-general case when the spin polarization that generates the $\vec{E}$-induced torque has components along all of the $X$, $Y$, and $Z$ axes, the allowed angular dependences for the coefficients $S$ and $A$ are[1,36,37]:

$$S = S_{DL}^Y \cos\phi \sin 2\phi + S_{DL}^X \sin\phi \sin 2\phi + S_{FL}^Z \sin 2\phi \tag{4}$$

$$A = A_{FL}^Y \cos\phi \sin 2\phi + A_{FL}^X \sin\phi \sin 2\phi + A_{DL}^Z \sin 2\phi \tag{5}$$

where $S_{DL}^Y$, $S_{DL}^X$, and $A_{DL}^Z$ are coefficients for the damping-like torque generated by the SHCs, $\sigma_{ZX}^Y$, $\sigma_{ZX}^X$, and $\sigma_{ZX}^Z$ respectively, while $A_{FL}^Y$, $A_{FL}^X$, and $S_{FL}^Z$ are their field-like-torque counterparts (and $A_{FL}^Y$ also contains the contribution from the Oersted torque). In our samples we can tell that the Oersted field is the dominant source for $A_{FL}^Y$ because there is good agreement between the measured strength of the out-of-plane field-like torque and the estimated contribution from the Oersted field due to the current flow through the RuO$_2$ layer (see Supplemental Material for more information). In this case, we can use the value of $A_{FL}^Y$ as a measure of the current density in the RuO$_2$, which allows us to quantify the amplitude of the damping-like torque efficiencies per unit current density in the RuO$_2$ ($j$) as:[36]

$$\xi_{DL,j}^X = \frac{S_{DL}^X}{A_{FL}^Y} \frac{e\mu_0 M_S t_{HM} t_{FM}}{\hbar} \sqrt{1 + \left(\frac{M_{eff}}{H_0}\right)} \tag{6}$$

$$\xi_{DL,j}^Y = \frac{S_{DL}^Y}{A_{FL}^Y} \frac{e\mu_0 M_S t_{HM} t_{FM}}{\hbar} \sqrt{1 + \left(\frac{M_{eff}}{H_0}\right)} \tag{7}$$

$$\xi_{DL,j}^Z = \frac{A_{DL}^Z}{A_{FL}^Y} \frac{e\mu_0 M_S t_{HM} t_{FM}}{\hbar} . \tag{8}$$

Here $\mu_0 M_{eff}$ is the magnetic anisotropy (= 0.77 ± 0.02 Tesla) measured by performing ST-FMR for a sequence of microwave frequencies and fitting the resonance fields to the Kittel equation. The saturation magnetization is $\mu_0 M_S$ = 0.78 ± 0.01 Tesla as measured by vibrating sample magnetometry, and $t_{HM}$ and $t_{FM}$ are the thicknesses of $RuO_2$ and Py. The damping-like torque efficiencies per unit applied electric field ($\xi_{DL,E}^k$) for each component of torque $k$ can be obtained by dividing by the resistivity $\rho_{xx}$ of the $RuO_2$ film,

$$\xi_{DL,E}^k = \xi_{DL,j}^k / \rho_{xx}. \tag{9}$$

For 6 nm thick $RuO_2$ at room temperature, $\rho_{xx}$ is approximately 140 μΩ-cm for (101)-oriented films and 275 μΩ-cm (001) films (compared to 75 μΩ-cm for the 4 nm thick Py). The components of the torque efficiencies per unit applied electric field $\xi_{DL,E}^k$ should be equal to the corresponding components of the spin Hall conductivity tensor $\sigma_{ZX}^k/[\hbar/(2e)]$ times an interfacial spin transparency that is less than or equal to 1.

For the in-plane HH method, we apply a low-frequency alternating current (1327 Hz) and measure both the first and second harmonic Hall voltages as a function of the angle of rotation $\phi$ of an in-plane magnetic field (Fig. 1(g)). The AC current generates alternating torques that result in oscillations in the Hall resistance due to the planar Hall effect and anomalous Hall effect. Mixing between the oscillating Hall resistance and the AC current results in a second-harmonic signal that can be expressed as[2,38,39]:

$$V_{XY}^{2\omega} = D_{DL}^Y \cos\phi + D_{DL}^X \sin\phi + D_{DL}^Z \cos 2\phi + F_{FL}^Y \cos\phi \cos 2\phi + F_{FL}^X \sin\phi \cos 2\phi + F_{FL}^Z \tag{10}$$

Damping-like torques generated by the *X*, *Y*, and *Z* component give rise to $D_{DL}^X$, $D_{DL}^Y$, and $D_{DL}^Z$, respectively, while the field-like torque counterparts give rise to $F_{FL}^X$, $F_{FL}^Y$ and $F_{FL}^Z$. The contributions from $\vec{E}$-induced torques are distinguished from thermoelectric voltages based on their dependence on the magnetic-field magnitude. See the Supplemental Material for the details of how the SHCs are calculated based on the HH measurements.

**Experimental results:**

Figure 2(b) shows the $\phi$-dependence of the ST-FMR amplitudes *S* (red squares) and *A* (blue circles) for a (001)-oriented $RuO_2$(6 nm)/Py(4 nm) sample when $\vec{E}$ is applied along the [001] axis. This is the crystal orientation for which $\mathcal{N}$ is oriented out-of-plane so that the AF-SHE should produce no unconventional $\vec{E}$-induced torque on the Py layer. We find that this is indeed the case. Both the symmetric component *S* and the antisymmetric component *A* can be fitted well simply using the conventional torque terms ∝ $\cos\phi \sin 2\phi$ that occur in ordinary heavy metals[36]. The strength of the symmetric term corresponds to a spin torque efficiency per unit electric field $\xi_{DL,E}^Y = \frac{\hbar}{2e}(1.8 \pm 0.4) \times 10^4$ $(\Omega m)^{-1}$ as determined from Eq. (7).

The ST-FMR results for a (101)-oriented RuO$_2$(6 nm)/Py (4 nm) sample, the geometry for which $\mathcal{N}$ is tilted away from the out-of-plane direction, are shown in Figs. 2(c-h). The data for $\vec{E}$ in the [010] direction ($\psi$=0° within alignment accuracy) are in Figs. 2(c,d). The symmetric component of the ST-FMR signal has an angular dependence similar to the (001) sample – only the conventional damping-like SHE term $S_{DL}^Y \cos\phi \sin 2\phi$ is required to fit the results (Fig. 2(c)), and the corresponding torque efficiency is $\xi_{DL,E}^Y = \frac{\hbar}{2e}(3.6 \pm 0.6) \times 10^4$ $(\Omega m)^{-1}$ (or $\xi_{DL,j}^y = 0.050 \pm 0.008$). Note that the antisymmetric component is very different than for the (001)-oriented film. For the (101)-oriented case, $A$ cannot be fit only by the $A_{FL}^Y \cos\phi \sin 2\phi$ term; there is also a significant $A_{DL}^Z \sin 2\phi$ contribution corresponding to an out-of-plane damping-like torque (with $\xi_{DL,E}^Z = (7 \pm 1) \times 10^3$ $(\Omega m)^{-1}$). This additional term is also consistent with our HH measurements (Fig. 2(g)) as $V_{XY}^{2\omega}$ requires the term $D_{DL}^Z \cos 2\phi$ for the fitting in addition to the conventional fitting function[2] $D_{DL}^Y \cos\phi + F_{FL}^Y \cos\phi \cos 2\phi$ (see Supplemental Material). In contrast, when $\vec{E}$ is applied along the [$\bar{1}$01] axis ($\psi$=90°) for the (101)-oriented case, we observe only conventional spin-torques, i.e. $\xi_{DL,E}^Y \neq 0, \xi_{DL,E}^X \approx 0, \xi_{DL,E}^Z \approx 0$, as both the $S$ and $A$ amplitudes in ST-FMR can be fit by $\cos\phi \sin 2\phi$ (Fig. 2(e,f)) and $V_{XY}^{2\omega}$ in the HH measurement can be fit by $D_{DL}^Y \cos\phi + F_{FL}^Y \cos\phi \cos 2\phi$ (Fig. 2(h)).

For the intermediate values of $\psi$ we find from ST-FMR measurements that the spin-torque efficiencies have the angular dependence shown in Fig. 3(a,b,c). The behaviors of all three components are in excellent agreement with Eqs. (1)-(3) for the fit parameters $C_0 = (26 \pm 1) \times 10^3$ $(\Omega m)^{-1}$, $C_1 = (14 \pm 2) \times 10^3$ $(\Omega m)^{-1}$, and $\theta_s = 44° \pm 5°$. This angular dependence is the most important result of this paper. It is a signature that the torques are indeed due to a spin current with a tilted spin orientation, with the spin flow direction perpendicular to both $\hat{s}$ and $\hat{E}$. To our knowledge, RuO$_2$ is the first material for which this complete signature angular dependence of an AF-SHE has been observed, corresponding to a $\vec{E}$-induced spin current with tilted spin. Furthermore, the orientation of $\hat{s}$ is close to the optimum angle for maximizing the out-of-plane damping-like torque, 45° according to Eq. (3).

We can identify the $\psi$-dependent torques as being generated by the antiferromagnetic order in RuO$_2$ by comparing to measurements on (101)-oriented IrO$_2$ films, which are isostructural with RuO$_2$, but non-magnetic. The (101)-oriented IrO$_2$ films generate only conventional spin-orbit torques, with a non-zero value of $\xi_{DL,E}^Y$, but with $\xi_{DL,E}^X = 0$ and $\xi_{DL,E}^Z = 0$ (Fig.4(a)). The lowered symmetry provided by the canted Néel vector of RuO$_2$ is therefore essential for the existence of the unconventional torque terms.

We mentioned above that the $\vec{E}$-induced spin $\hat{s}$ is expected to be only approximately parallel to the Néel vector $\mathcal{N}$, as shown in Fig. 1. This is not expected to be exact because deviation from the Néel vector orientation can be generated by SOI. Assuming $\mathcal{N}$ is oriented along the [001] direction of the RuO$_2$ crystal, and given the lattice constants of bulk RuO$_2$ (0.312 nm in the [001] direction and 0.459 nm in the perpendicular directions), in a (101)-oriented film the tilt angle of $\mathcal{N}$ relative to the out-of-plane direction should be about 34°, a smaller tilt angle than for the spin orientation $\hat{s}$ indicated by the angular dependence in Fig. 3. This difference is qualitatively consistent with our density functional theory calculations, suggesting that for the AF-SHE, $\hat{s}$ is tilted from $\mathcal{N}$ for a (101)-oriented film by about 5° within the (010) plane due to SOI, which generates additional spin current components non-collinear to the Néel vector (see Supplemental Material for details). Other possible contributions that might alter $\theta_s$ slightly are that $\mathcal{N}$ might not be oriented strictly along the [001] direction as reported in ref. 30, epitaxial strain can modify the lattice parameters slightly, and the interface spin transmission factor could depend on the orientation of the spin, so that spin absorbed by the ferromagnetic layer to generate the $\vec{E}$-induced torque might not be exactly aligned with the spin orientation of the spin current within the RuO$_2$ layer.

We measured the damping-like $\vec{E}$-induced torques for different thicknesses of the (101)-oriented RuO$_2$ films (3-9 nm) as shown in Fig. 4(b) for measurements with $\psi = 0°$ (squares) and $\psi = 90°$ (open circles). Both the conventional ($\psi$-independent) and unconventional ($\psi$-dependent) torque efficiencies increase with increasing RuO$_2$ thickness, suggesting that the $\vec{E}$-induced spin current originates in the bulk of the RuO$_2$ film rather than at the RuO$_2$/Py interface. The approach to saturation at larger thicknesses suggests a spin diffusion length[36] of 2.6 ± 0.3 nm in the RuO$_2$. Additional evidence for a bulk generation mechanism comes from samples in which we inserted a thin Ir spacer between the RuO$_2$ and the Py (the full structure was RuO$_2$(9 nm)/Ir(1.2 nm)/Py(4 nm)/Ir(1.2 nm)). In this case we still observe non-zero values of both $\xi^y_{DL,E} = (3.5 \pm 0.7) \times 10^4$ $(\Omega m)^{-1}$ and $\xi^z_{DL,E} = (3.3 \pm 0.6) \times 10^3$ $(\Omega m)^{-1}$, only slightly reduced compared to RuO$_2$(9 nm)/Py(4 nm) with no spacer (Fig. 4(b)).

Based on the evidence we have presented so far, we have not determined whether the unconventional $\vec{E}$-induced torques generated by RuO$_2$ correspond to a $\mathcal{T}$-odd or a $\mathcal{T}$-even spin current, because the AF order in RuO$_2$ allows both effects, and they could both generate spin currents with a well-defined orientation $\hat{s}$, giving a qualitatively-similar dependence on the angle $\psi$ (see Supplemental Material for more information). Our measured SHCs are approximately a factor of 50 smaller than the prediction for the $\mathcal{T}$-odd spin current from a uniform single-domain RuO$_2$ sample (Fig. 1(d) and Ref. 4), but this could be due to nearly-complete cancellation between domains with $\mathcal{N}$ in the [001] and [00$\bar{1}$] directions. Based just on the amplitude of the torques, we also cannot rule out a contribution from the $\mathcal{T}$-even spin current even though that is much smaller than the $\mathcal{T}$-odd spin current, since the predicted value is similar to the order of magnitude that we report here.

To help distinguish the $\mathcal{T}$-odd and $\mathcal{T}$-even mechanisms, we have measured the out-of-plane damping-like spin-Hall conductivity as a function of temperature (Fig. 4(d)). We observe a strong enhancement as a function of decreasing temperature. That is suggestive that the $\mathcal{T}$-odd spin current is dominant, since this should be enhanced as the electron lifetime increases at low temperature (see Supplemental Material). Conventional $\mathcal{T}$-even SHCs arising from intrinsic spin Hall effects generally have negligible temperature dependence for moderately-low-resistivity materials like (101) RuO$_2$[28]. We observe a non-zero out-of-plane damping-like torque up to at least 400 K (Fig. 4(d)), confirming that AF ordering in RuO$_2$ persists up to this temperature[29,30].

Even with the likelihood that the strength of the out-of-plane damping-like torque is reduced by cancellations between domain variants, the strength of the effect is still among the largest of all materials that have been measured. For 6 nm thick (101)-oriented RuO$_2$, from Fig. 3(b) we measure a maximum room-temperature value for the out-of-plane damping-like spin torque efficiency per unit electric field of $\xi^Z_{DL,E} = (7 \pm 1) \times 10^3$ $(\Omega m)^{-1}$, compared to, e.g., $(3.6 \pm 0.8) \times 10^3$ $(\Omega m)^{-1}$ for WTe$_2$[1] and $(1.02 \pm 0.03) \times 10^3$ $(\Omega m)^{-1}$ for MoTe$_2$[40]. The only materials for which we are aware of a slightly larger out-of-plane damping-like spin-torque efficiency are the non-collinear antiferromagnet Mn$_3$GaN[20] with $\xi^Z_{DL,E} \approx 8.6 \times 10^3$ $(\Omega m)^{-1}$ and (114)-textured MnPd$_3$[22] with $\xi^Z_{DL,E} \approx 1.4 \times 10^4$ $(\Omega m)^{-1}$. If our results are indeed due to a $\mathcal{T}$-odd spin current, the strength of the out-of-plane damping-like torque in RuO$_2$ might be increased substantially by controlling the AF domains to more-strongly favor one of the domain variants ($\mathcal{N}$ parallel to [001] or [00$\bar{1}$]) over the other.

**Conclusion**

We have shown that (101)-oriented films of the collinear antiferromagnet $RuO_2$ produce a significant electric-field ($\vec{E}$) generated out-of-plane damping-like torque on an adjacent Permalloy film. Detailed measurements of the dependence of this torque as a function of the angles of both $\vec{E}$ and magnetic field relative to the crystal axes demonstrate that the torque is associated with a spin polarization $\hat{s}$ approximately parallel to the Néel vector of the $RuO_2$, with a spin flow direction perpendicular to both $\hat{s}$ and $\vec{E}$. This angular dependence is the signature of an antiferromagnetic spin Hall effect predicted for $RuO_2$[23].

**Acknowledgements**

We thank Thow Min Cham for discussions. A.B. played the primary role in sample fabrication, and together with R.J. led the spin-torque measurements and data analysis, supervised by D.C.R. R.J. was supported by the US Department of Energy (DE-SC0017671) and A.B was supported in part by the DOE and in part by the Cornell Center for Materials Research (supported by the National Science Foundation (NSF) MRSEC program, DMR-1719875). N.J.S., H.P.N., and J.S. grew the $RuO_2$ and $IrO_2$ films, supervised by D.G.S. N.J.S., H.P.N., J. S., and D.G.S. acknowledge support from the NSF Platform for the Accelerated Realization, Analysis and Discovery of Interface Materials (PARADIM) under Cooperative Agreement No. 2039380, and this research is funded in part by the Gordon and Betty Moore Foundation's EPiQS Initiative, Grant GBMF9073 to D.G.S. D-F.S. performed the DFT calculations, supported by the NSF MRSEC program (DMR-1420645) and supervised by E.Y.T. X.Z. performed STEM imaging, supported by the NSF MRSEC program (DMR-1719875) and the NSF MRI program (DMR-1429155) and supervised by D.A.M. The devices were fabricated using the shared facilities of the Cornell NanoScale Facility, a member of the National Nanotechnology Coordinated Infrastructure (supported by the NSF, NNCI-2025233) and the facilities of Cornell Center for Materials Research.

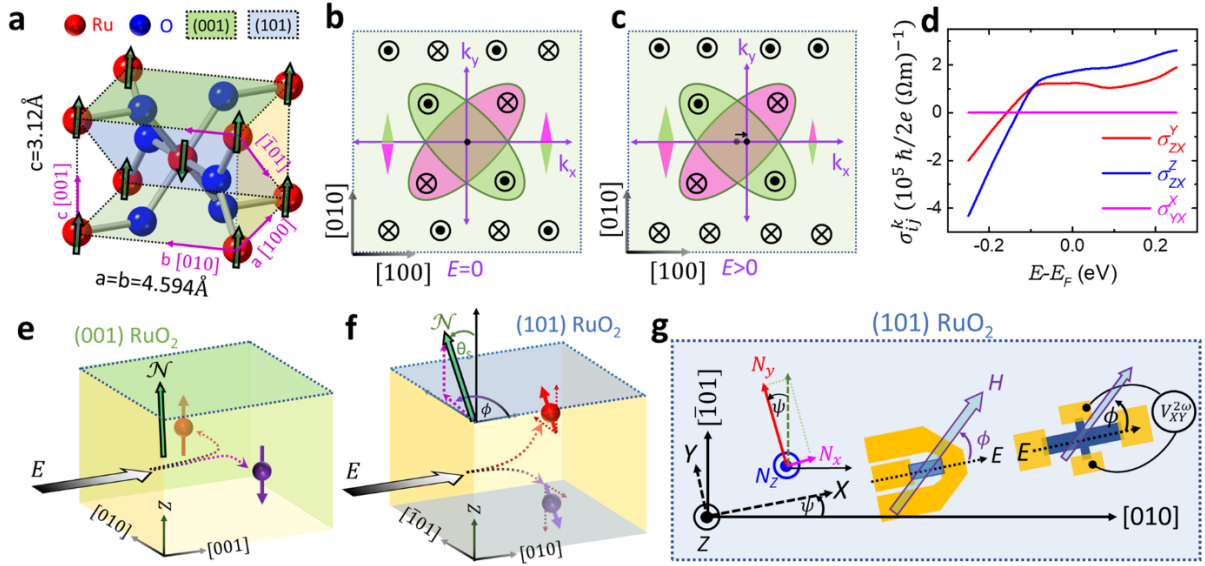

Fig. 1.

(a) Crystal structure of RuO$_2$ with spin orientations indicated (green arrows). (b,c) Schematic diagrams of the spin-split Fermi surface of RuO$_2$ for up-spin and down-spin electrons with (b) zero and (c) nonzero applied electric field in the [001] direction. The green and pink triangles denote the transverse flow of up and down-spin electrons. (d) Calculated time-odd spin Hall conductivity for a (101)-oriented RuO$_2$ film with applied electric field in the [010] direction. (e,f) Illustration of spin current generation due to the antiferromagnetic spin Hall effect for (e) a (001)-oriented RuO$_2$ film and (f) a (101) film. (g) Sample schematics and definition of coordinate axes for the (101)-oriented sample. The $X$ direction is parallel to the applied electric field and $\psi$ is the angle of the electric field relative to the [010] crystal axis.

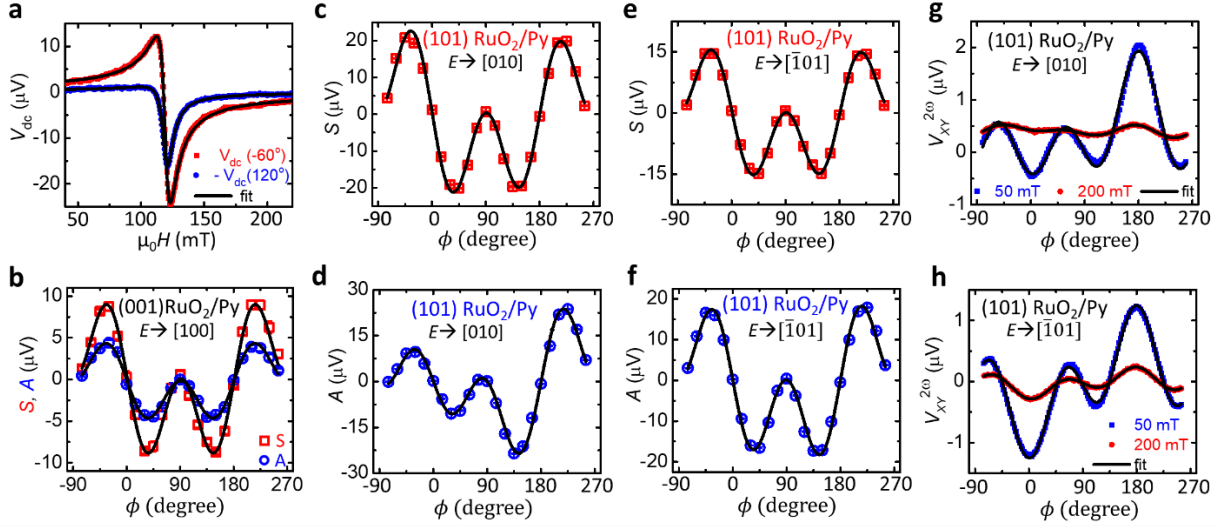

Fig. 2.

(a) ST-FMR signals measured for a (101)-oriented RuO$_2$(6 nm)/Py(4 nm) sample with electric field applied along the [010] axis with a 15 dbm, 10 GHz microwave drive at room temperature, along with fits to the sum of symmetric and antisymmetric line shapes. The two curves shown correspond to magnetic-field sweeps along the $\phi = -60°$ and $\phi = 120°$ directions. (b) Symmetric resonance amplitude (S) vs. in-plane magnetic-field angle ($\phi$) (red squares) and antisymmetric resonance amplitude (A) vs. $\phi$ (blue circles) for a (001)RuO$_2$(6 nm)/Py(4 nm) sample with electric field applied along the [100] axis. (c) S vs. $\phi$ and (d) A vs. $\phi$ for a (101)RuO$_2$(6 nm)/Py(4 nm) sample with electric field applied along the [010] axis ($\psi = 0°$). (e) S vs. $\phi$ and (f) A vs. $\phi$ for a (101)RuO$_2$(6 nm)/Py(4 nm) sample with electric field applied along the [$\bar{1}$01] axis ($\psi = 90°$). Harmonic Hall data for a (101)RuO$_2$(6 nm)/Py(4 nm) sample with electric field applied along (g) the [010] direction and (h) the [$\bar{1}$01] direction for two different magnitudes of in-plane magnetic field.

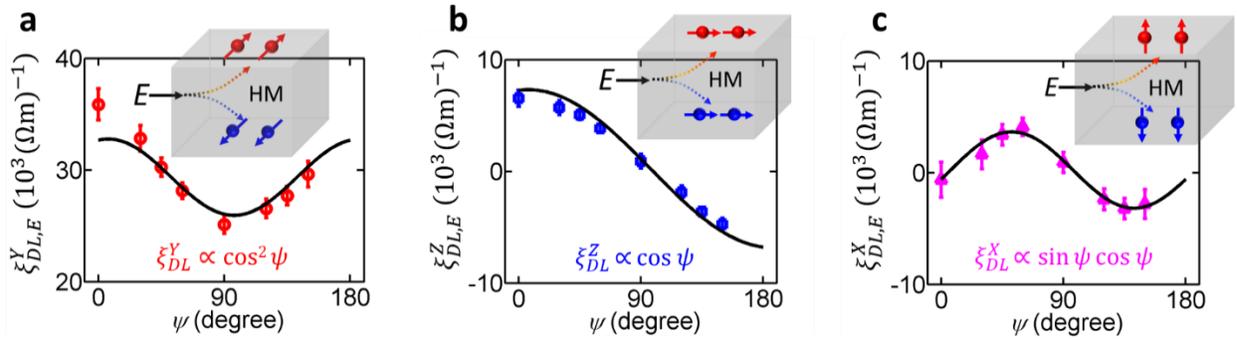

Fig. 3.

For (101)RuO$_2$(6 nm)/Py(4 nm) samples, the dependence of the measured components of the damping-like torque efficiency per unit applied electric field on the angle ($\psi$) of the electric field relative to the [010] direction, corresponding to the (a) $\hat{Y}$ component of the spin current, (b) the $\hat{Z}$ component, and (c) the $\hat{X}$ component.

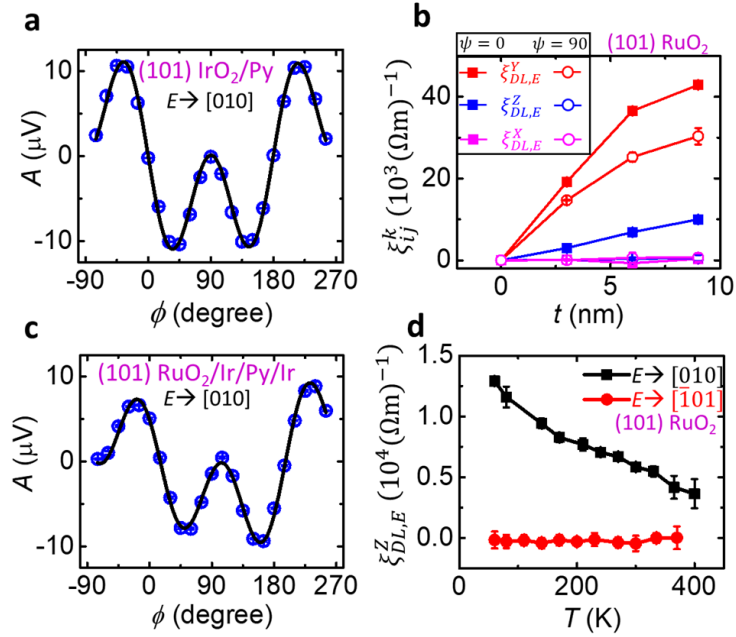

Fig. 4.

(a) $A$ vs. $\phi$ for a (101)IrO$_2$(6 nm)/Py(4 nm) sample with electric field applied along the [010] direction ($\psi = 0°$), showing the absence of any unconventional torques from non-magnetic (101)IrO$_2$. (b) Measured spin torque efficiencies per unit applied electric field in (101)RuO$_2$/Py(4 nm) samples as a function of RuO$_2$ thickness for $\psi = 0°$ (filled squares) and $\psi = 90°$ (open circles). (c) $A$ vs. $\phi$ for a (101)RuO$_2$(9 nm)/Ir(1 nm)/Py(4 nm)/Ir(1 nm) sample with electric field applied along the [010] direction ($\psi = 0°$), showing the persistence of unconventional torques even with an Ir spacer layer. (d) Out-of-plane damping-like spin-torque efficiency per unit electric field for a (101)RuO$_2$(6 nm)/Py(4 nm) sample as a function of temperature for $\psi = 0°$ (black squares) and $\psi = 90°$ (red circles).

# Supplemental Material

## Tilted spin current generated by the collinear antiferromagnet RuO$_2$

Arnab Bose*[1,2], Nathaniel J. Schreiber*[3], Rakshit Jain*[1,2], Ding-Fu Shao[4], Hari P. Nair[3], Jiaxin Sun[3], Xiyue Zhang[1], David A. Muller[1,5], Evgeny Y. Tsymbal[4], Darrell G. Schlom[3,5], and Daniel C. Ralph[2,5]

1 School of Applied and Engineering Physics, Cornell University. Ithaca, NY 14853, USA

2 Department of Physics, Cornell University. Ithaca, NY 14853, USA

3 Department of Materials Science and Engineering, Cornell University. Ithaca, NY 14853, USA

4 Department of Physics and Astronomy & Nebraska Center for Materials and Nanoscience, University of Nebraska, Lincoln, Nebraska 68588-0299, USA

5 Kavli Institute at Cornell for NanoScale Science, Ithaca, NY 14830, USA

* These authors contributed equally

**S1. Details of device preparation**

**S2. Details of the second-harmonic Hall analysis**

**S3. Details of the ST-FMR analysis**

**S4. Calculation of spin Hall conductivities**

**S5. Exchange bias in (101)RuO$_2$/Py samples**

**S1. Details of device preparation**

The RuO$_2$ thin films were prepared by reactive-oxide molecular beam epitaxy (MBE) in a background pressure of $3 \times 10^{-6}$ Torr of distilled ozone (~80% O$_3$ + 20% O$_2$) at a substrate temperature of 310 °C. A Ru atomic flux of $6 \times 10^{12}$ atoms cm$^{-2}$ s$^{-1}$ was supplied by an electron beam evaporator. Companion RuO$_2$ films were characterized by both θ-2θ X-ray diffraction (XRD) and X-ray reflectivity (XRR) performed with a Rigaku Smartlab high-resolution diffractometer using Cu-Kα1 radiation (λ = 1.5406 Å) and a Ge(220) double-bounce monochromator on the incident-beam side (Fig. S1(a)). The films were also examined by cross-sectional scanning transmission electron microscopy (STEM) on samples prepared using a Thermo Helios G4 UX Focused Ion Beam with a final milling step of 5 keV to reduce surface damage. High-resolution high-angle annular dark field STEM images were acquired using an aberration-corrected 300-keV Themis Titan microscope. These images indicate an atomically-sharp interface between single-crystal (101)-oriented RuO$_2$ and the polycrystalline Py overlayer (Fig. S1(b)). Further details of RuO$_2$ growth and thin film characterization can be found in ref. [1].

After the RuO$_2$ growth, the films were transported in air to a sputtering system with a base vacuum better than $2 \times 10^{-8}$ Torr, and 4 nm of Py was deposited at 30 mW power and 2 mTorr Ar pressure. To protect the Py layer from oxidation, we capped the Py with 1.2 nm Ta which naturally oxidizes to TaO$_X$. Devices were then patterned by optical lithography and argon-ion milling such that electric current could be applied at different angles ($\psi$) with respect to the crystal axis as shown in Fig. S1(c). Electrical contacts of Ti(5 nm)/Pt(75 nm) were made by optical lithography, sputtering, and lift-off. For the spin-torque ferromagnetic resonance (ST-FMR) studies the electrical contacts were in the form of a microwave-compatible ground-source-ground geometry (Fig. S1(c)), and the dimensions of the ST-FMR device under test were 30 μm long × 20 μm wide. For the Harmonic Hall (HH) measurements the length and the width of the Hall bars were 20 μm and 6 μm.

The resistivity of (101)-$RuO_2$ and the polycrystalline Py films are approximately 140 μΩ-cm and 70 μΩ-cm respectively as determined by 4-probe resistance measurements. (The resistivity of (101) $RuO_2$ has negligible dependence as a function of the angle $\psi$.) We define the coefficients for the planar Hall effect ($R_P$) and anomalous Hall effect ($R_A$) as:

$$R_{XY} = R_0 + R_P \sin 2\phi \sin^2\theta + R_A \cos\theta$$

Fig. S2(a,b) shows measurements of $R_{XY}$ for a (101)$RuO_2$(6 nm)/Py(4 nm) sample. Permalloy has a strong planar Hall effect and a weak anomalous Hall effect compared to Co or CoFeB [2]. (The strong planar Hall effect causes the kinks that are visible at low field values in Fig. S2(a).) The out-of-plane demagnetization field ($\mu_0 H_K$) of 4 nm thick Py is approximately 0.78 T as obtained from the out-of-plane field sweep (Fig. S2(a)). This value is close to the bulk value for Py.

**S2. Details of the second-harmonic Hall analysis**

For second-harmonic Hall measurements on (101)-oriented $RuO_2$/Py samples, the current axis of the Hall bar for different devices is aligned at various angles relative to the [010] crystal axis (Fig. S1): $\psi$ = 0°, 30°, 45°, 60°, 90°, 120°, 135° and 150°. We apply a low-frequency alternating current $I(t) = \Delta I \cos(2\pi f t)$ (with $f$ = 1327 Hz) and an in-plane magnetic field $H_{ext}$ at various angles $\phi$ relative to the current-flow direction, and measure the first and second-harmonic voltages using a lock-in amplifier. We investigated fixed values of $\mu_0 H_{ext}$ ranging from 50 to 300 mT, applied by a GMW projected-field magnet on a Newport motion-controlled stage.

The alternating current through $RuO_2$ produces both in-plane and out-of plane torques on the Py layer, causing deflections in the magnetic orientation that change both the anomalous Hall effect and the planar Hall effect. Mixing with the alternating current then generates a second-harmonic Hall voltage that can be expressed [3–6]:

$$V_{2\omega} = D_{DL}^Y \cos\phi + D_{DL}^X \sin\phi + D_{DL}^Z \cos 2\phi + F_{FL}^Y \cos\phi \cos 2\phi + F_{FL}^X \sin\phi \cos 2\phi + F_{FL}^Z \quad (1)$$

with

$$D_{DL}^Y = -\frac{\tau_{DL,0}^Y}{\gamma} \frac{V_A}{2(H_{ext}+H_k)} + V_{ANE} + V_{ONE} H_{ext} \quad (2)$$

$$D_{DL}^X = -\frac{\tau_{DL,0}^X}{\gamma} \frac{V_A}{2(H_{ext}+H_k)} \quad (3)$$

$$D_{DL}^Z = -\frac{\tau_{DL,0}^Z}{\gamma} \frac{V_P}{H_{ext}} \quad (4)$$

$$F_{FL}^Y = -H_{FL}^Y \frac{V_P}{H_{ext}} \quad (5)$$

$$F_{FL}^X = -H_{FL}^X \frac{V_P}{H_{ext}} \quad (6)$$

$$F_{FL}^Z = H_{FL}^Z \frac{V_A}{2(H_{ext}+H_k)} + C. \tag{7}$$

Here, the damping-like torques ($\vec{\tau} \propto d\hat{m}/dt$) produced by spins in the X, Y, and Z directions are $\vec{\tau}_{DL,X} = \tau_{DL,0}^X \hat{m} \times (\hat{X} \times \hat{m})$, $\vec{\tau}_{DL,Y} = \tau_{DL,0}^Y \hat{m} \times (-\hat{Y} \times \hat{m})$, and $\vec{\tau}_{DL,Z} = \tau_{DL,0}^Z \hat{m} \times (\hat{Z} \times \hat{m})$, and $\vec{H}_{FL,X} = H_{FL}^x \hat{X}$, $\vec{H}_{FL,Y} = -H_{FL}^y \hat{Y}$, and $\vec{H}_{FL,Z} = H_{FL}^z \hat{z}$ are effective fields associated with the field-like torques. (The in-plane Oersted field as well as spin-transfer torques can contribute to $H_{FL}^Y$, and the signs of the Y-components are picked so that $\tau_{DL,0}^Y$ is positive for Pt, and $H_{FL}^Y$ is positive for an Oersted field.) $V_P$ is the coefficient of the peak planar Hall effect voltage, $V_{PHE} = V_P \sin 2\phi$, for the given amplitude of alternating current, and $V_A$ is the coefficient of the peak anomalous Hall voltage $V_{AHE} = V_A \cos \theta$. $V_{ANE}$ is an anomalous Nernst voltage, $V_{ONE} H_{ext}$ is a voltage due to the ordinary Nernst effect, $H_k$ characterizes the out-of-plane magnetic anisotropy (positive for samples with in-plane anisotropy), and C accounts for a constant background. The damping-like torque efficiencies per unit electric field associated with each component $k$ are

$$\xi_{DL,E}^k = \frac{2e}{\hbar} \frac{\mu_0 M_S t_{FM}}{\gamma} \frac{\tau_{DL,0}^k}{E}, \tag{8}$$

and the corresponding field-like torque efficiencies per unit electric field are

$$\xi_{FL,E}^k = \frac{2e}{\hbar} M_S t_{FM} \frac{\mu_0 H_{FL}^k}{E}. \tag{9}$$

Here $M_S$ is the saturation magnetization, $E$ is the peak electric field associated with the alternating current, $e$ is electronic charge, $\hbar$ is Plank's constant and $t_{FM}$ is the thickness of ferromagnet (Py here).

Because for our samples the value of $\mu_0 H_k \approx 0.78$ T is large and the anomalous Hall voltage $V_A$ is relatively small, the contributions to $V_{2\omega}$ from the out-of-plane effective fields (meaning the in-plane torques, Eqs. (2), (3), and (7)) is small, making these components difficult to measure accurately by this technique. For the harmonic-Hall measurements, we will therefore focus on the in-plane effective fields (*i.e.*, out-of-plane torques) that can be measured accurately, the terms containing: $\tau_{DL,0}^Z$, $H_{FL}^Y$, and $H_{FL}^X$.

Examples of the $\phi$ dependence for the second-harmonic Hall voltage of (101)RuO$_2$(6 nm)/Py(4 nm) samples are shown in Fig. S2(c) for electric field applied in the [010] direction ($\psi = 0°$) and in Fig. S2(d) for electric fields in the [$\bar{1}$01] direction ($\psi = 90°$). In each figure, we show scans for two values of magnetic-field magnitude, 50 mT (red) and 300 mT (blue). Fits to Eq. (1) are shown in black lines.

To quantify the effective fields for the out-of plane torque terms, $F_{FL}^Y$, $D_{DL}^Z$ and $F_{FL}^X$ are normalized with respect to $V_P$, plotted as a function of $1/(\mu_0 H_{ext})$, and fit to straight lines (Fig. S2(e-g) for $\psi=0°$ and Fig. S3(h-j) for $\psi=90°$). From the slopes of the fit lines and Eqs. (8) and (9) we calculate the torque efficiencies $\xi_{FL,E}^Y$, $\xi_{DL,E}^Z$, and $\xi_{FL,E}^X$.

The field-like efficiency $\xi_{FL,E}^Y$ (Fig. S2(k)) is approximately constant, with only a small variation with $\psi$. This component is dominated simply by the current-induced Oersted field, $H_{Oe} = \frac{1}{2} J_{HM} t_{HM}$, where $J_{HM}$ is the current density within the RuO$_2$ layer. Using Eq. (9), the Oersted field translates into a torque efficiency

$$\xi_{Oersted,E}^Y = \frac{2e}{\hbar} \frac{\mu_0 M_S t_{FM} t_{HM}}{2\rho_{xx}}, \tag{10}$$

where $\rho_{xx}$ is the resistivity of the RuO$_2$ layer. Substituting in the values appropriate for our samples, $\mu_0 M_S = 0.78 \pm 0.01$, $t_{FM} = 4$ nm, $t_{HM} = 6$ nm, and $\rho_{xx} = 140$ μΩ-cm yields the blue dashed line in Fig. S2(k). We conclude that the Oersted torque contributes at least 90% of the in-plane field-like torque efficiency $\xi_{FL,E}^Y$ regardless of the electric-field angle $\psi$.

The out-of-plane damping-like torque efficiency $\xi_{DL,E}^Z$ extracted from the second-harmonic Hall measurements is shown in Fig. S2(l). Here there is a strong dependence on $\psi$, confirming the results of the ST-FMR measurement shown in Fig. 3(b) of the main text. The harmonic-Hall and ST-FMR measurements are also in reasonable quantitative agreement, with a maximum value determined from the harmonic-Hall measurement being $\xi_{DL,E}^Z = (7.5 \pm 1.0) \times 10^3$ $(\Omega m)^{-1}$, compared to $\xi_{DL,E}^Z = (7 \pm 1) \times 10^3$ $(\Omega m)^{-1}$ for the ST-FMR results described in the main text.

The results for the torque efficiency $\xi_{FL,E}^X$ are shown in Fig. S2(m). This term is close to zero, making it difficult to extract any meaningful dependence on $\psi$.

### S3. Details of the ST-FMR analysis

For the ST-FMR measurements, we also pattern different devices so that the applied microwave current will have different angles, $\psi$, relative to the crystallographic axes as shown in Fig. S1(a). For each device, an in-plane magnetic field is swept from 2700 Oe to 300 Oe at a variety of azimuthal angles, $\phi$, with respect to the applied microwave-current direction using a GMW projected-field magnet on a Newport motion-controlled stage. We apply microwave current amplitude-modulated at a reference frequency of 1327 Hz using an Agilent current source, and measure the ST-FMR mixing voltage using a lock-in amplifier.

We found in the previous section based on harmonic-Hall measurements that the out-of-plane field-like torque efficiency $\xi_{FL,E}^Y$ agreed with the value expected solely from the Oersted field within 10%. From this we infer that the corresponding out-of-plane field-like-torque term $A_{FL}^Y$ in the ST-FMR measurements should be similarly dominated by the Oersted field. We checked this directly by calibrating the microwave current in selected samples using a vector network analyzer [7,8], and from this estimating the current density $J_{HM}$ flowing within the RuO$_2$ layer within a parallel-conductor model. The Oersted torque based on this estimate, $\gamma \mu_0 J_{HM} t_{HM}/2$ also agrees with the measured out-of-plane field-like torque in the ST-FMR measurements to within 10%.

Because the out-of-plane field-like torque is dominated by the Oersted field, we can use this out-of-plane field-like term as a measure of the current density within the RuO$_2$ layer, $J_{HM} = 2A_{FL}^Y/(\gamma t_{HM})$. Equations (6)-(8) in the main text then follow, and provide a way to determine the various components of the spin-torque efficiency as ratios of the coefficients determined by fits to Eqs. (4) and (5) in the main text [9].

Figure S3 shows the symmetric (*S*) and antisymmetric (*A*) ST-FMR amplitudes as a function of the magnetic field angle $\phi$ for (101)-oriented RuO$_2$(6 nm)/Py(4 nm) samples with different current orientations $\psi$ relative to the [010] axis, along with fits to Eqs. (4) and (5) in the main text. For $\psi = 0°$ and 90° we find that *S* can be nicely fit by a $\cos\phi \sin 2\phi$ term alone (Fig. S3(a,c)), indicating a dominant contribution from the conventional damping-like torque ($\xi_{DL,E}^Y$). In contrast for intermediate angles of $\psi$ the fits to *S* require an additional $\sin\phi \sin 2\phi$ component (Fig. S3(b,d)) that changes sign between $\psi = 45°$ and $\psi = 135°$. This reflects the variation of $\xi_{DL,E}^X$ proportional to $\sin 2\psi$ as discussed for Fig. 3(c) in the main text.

We find that for $\psi = 90°$ the antisymmetric component *A* can be fit well using only a $\cos\phi \sin 2\phi$ term, indicating a conventional origin for the out-of-plane field like torque at this angle (dominated by the Oersted field as discussed above). In contrast for $\psi = 0°$ fits to $A$ require an additional $\sin 2\phi$ factor, indicating the presence of a strong out-of-plane damping-like torque efficiency $\xi_{DL,E}^Z$. As we tune from $\psi = 0°$ to $\psi = 90°$ this out-of-plane damping-like torque gradually decreases (Fig. S3(e-g)) and then changes sign for $\psi > 90°$ (Fig. S3(h)). This angular dependence is proportional to $\cos \psi$ as shown in Fig. 3(b) in the main text.

In addition to the damping-like torque components shown in Fig. 3 in the main text, ST-FMR also allows detection of the field-like torque components (Fig. S3(i-k)). The conventional term $\xi_{FL,E}^Y$ is in good agreement with the contribution expected from the Oersted field (with the microwave current calibrated by a vector network analyzer measurement), and the unconventional terms $\xi_{FL,E}^Z$ and $\xi_{FL,E}^X$ are zero within experimental accuracy with no apparent dependence on $\psi$.

## S4. Calculation of spin Hall conductivities

We calculate the atomic and electronic structures of RuO$_2$ using the projector augmented-wave (PAW) method [10] implemented in the VASP code [11]. A plane-wave cut-off energy of 500 eV and a 16 × 16 × 16 $\vec{k}$-point mesh in the irreducible Brillouin zone are used in the calculations. The exchange and correlation effects are treated within the generalized gradient approximation (GGA) developed by Perdew-Burke-Ernzerhof (PBE) [12]. The GGA+U functional [13,14] with $U_{\text{eff}} = 2$ eV on Ru 4$d$ orbitals and $U_{\text{eff}} = 5$ eV on Ti 3$d$ orbitals is included in all the calculations. Spin-orbit interaction (SOI) is included in all the calculations. We used the tight-binding Hamiltonians obtained from the maximally-localized Wannier functions [15] within the Wannier90 code [16]. The time-reversal odd and even parts of spin Hall conductivity (SHC) are given by [17,18]

$$\sigma_{ij}^k = -\frac{e\hbar}{\pi} \int \frac{d^3 \vec{k}}{(2\pi)^3} \sum_{n,m} \frac{\Gamma^2 \text{Re}(\langle n\vec{k}|J_i^k|m\vec{k}\rangle\langle m\vec{k}|v_j|n\vec{k}\rangle)}{\left[(E_F - E_{n\vec{k}})^2 + \Gamma^2\right]\left[(E_F - E_{m\vec{k}})^2 + \Gamma^2\right]}, \quad (11)$$

and

$$\sigma_{ij}^k = -\frac{2e}{\hbar} \int \frac{d^3 \vec{k}}{(2\pi)^3} \sum_{n' \neq n} \frac{\text{Im}(\langle n\vec{k}|J_i^k|n'\vec{k}\rangle\langle n'\vec{k}|v_j|n\vec{k}\rangle)}{\left(E_{n\vec{k}} - E_{n'\vec{k}}\right)^2}, \quad (12)$$

where $J_i^k = \frac{1}{2}\{v_i, s_k\}$ is the spin-current operator, $f_{n\vec{k}}$ is the Fermi-Dirac distribution function for band $n$ and wave vector $\vec{k}$, $v_i$ and $s_k$ are velocity and spin operators, respectively, and $i, j, k = x, y, z$. A 500 × 500 × 500 $k$-point mesh was used for the integral of Eqs. (11) and (12). When calculating the $\mathcal{T}$-odd SHC in Eq. (11), a constant $\Gamma$ that determines the broadening magnitude was used, which can be estimated by comparing the calculated conductivity with the experimental conductivity. For RuO$_2$ the room temperature conductivities are 3600 $\Omega^{-1}\text{cm}^{-1}$ for (001) film and 6900 $\Omega^{-1}\text{cm}^{-1}$ for (101) film, corresponding to $\Gamma \approx 50$ meV and $\Gamma \approx 25$ meV, respectively. When calculating the $\mathcal{T}$-even SHC in Eq. (12), the adaptive smearing method [19] was used.

Table S1 shows the calculated spin Hall conductivity tensors in units of $10^5 \left(\frac{\hbar}{2e}\right)(\Omega\, m)^{-1}$ for RuO$_2$ films with different growth orientations and charge current directions. For a (001) film, only conventional spin Hall conductivities $\sigma_{ij}^k$ ($i \neq j \neq k$) are non-zero. For a (101) film, the coordinate system is transformed from that of a (001) film using a rotation matrix $D$. The spin Hall conductivity tensor of RuO$_2$ (101) film is then obtained as follows $\sigma_{(101)_{i,j}}^{s,k} = \sum_{l,m,n} D_{ij} D_{jm} D_{kn}\, \sigma_{(001)_{lm}}^{s,n}$.

For the (101) oriented film we list in Table S1 values for the SHC tensors for *x* and *y* along the [010] and [$\bar{1}$01] directions. The $\mathcal{T}$-even and $\mathcal{T}$-odd SHCs for spin flow in the out-of-plane direction as a function of $\psi$ are shown in Fig. S4(a) and S4(b) respectively.

Note the SOI is included in all the calculations of the SHC. In the absence of SOI, some of the $\mathcal{T}$-odd components and all the $\mathcal{T}$-even components of SHC are vanishing (denoted by the grey background in Table S1).

## Calculated SHC of RuO$_2$ (001) and (101) films

| | | $\sigma^x$ $\begin{bmatrix} \sigma_{xx}^{x0} & \sigma_{xy}^{x0} & \sigma_{xz}^{x0} \\ \sigma_{yx}^{x0} & \sigma_{yy}^{x0} & \sigma_{yz}^{x0} \\ \sigma_{zx}^{x0} & \sigma_{zy}^{x0} & \sigma_{zz}^{x0} \end{bmatrix}$ | $\sigma^y$ $\begin{bmatrix} \sigma_{xx}^{y0} & \sigma_{xy}^{y0} & \sigma_{xz}^{y0} \\ \sigma_{yx}^{y0} & \sigma_{yy}^{y0} & \sigma_{yz}^{y0} \\ \sigma_{zx}^{y0} & \sigma_{zy}^{y0} & \sigma_{zz}^{y0} \end{bmatrix}$ | $\sigma^z$ $\begin{bmatrix} \sigma_{xx}^{z0} & \sigma_{xy}^{z0} & \sigma_{xz}^{z0} \\ \sigma_{yx}^{z0} & \sigma_{yy}^{z0} & \sigma_{yz}^{z0} \\ \sigma_{zx}^{z0} & \sigma_{zy}^{z0} & \sigma_{zz}^{z0} \end{bmatrix}$ |
|---|---|---|---|---|
| RuO$_2$ (001) *x*[100] *y*[010] | $\mathcal{T}$-odd $\Gamma \approx 25$ meV | $\begin{bmatrix} 0 & 0 & 0 \\ 0 & 0 & A=0.8273 \\ 0 & B=-0.0461 & 0 \end{bmatrix}$ | $\begin{bmatrix} 0 & 0 & A=0.8273 \\ 0 & 0 & 0 \\ B=-0.0461 & 0 & 0 \end{bmatrix}$ | $\begin{bmatrix} 0 & C=3.7452 & 0 \\ C=3.7452 & 0 & 0 \\ 0 & 0 & 0 \end{bmatrix}$ |
| | $\mathcal{T}$-odd $\Gamma \approx 50$ meV | $\begin{bmatrix} 0 & 0 & 0 \\ 0 & 0 & 0.4252 \\ 0 & -0.0150 & 0 \end{bmatrix}$ | $\begin{bmatrix} 0 & 0 & 0.4252 \\ 0 & 0 & 0 \\ -0.0150 & 0 & 0 \end{bmatrix}$ | $\begin{bmatrix} 0 & 1.9156 & 0 \\ 1.9156 & 0 & 0 \\ 0 & 0 & 0 \end{bmatrix}$ |
| | $\mathcal{T}$-even | $\begin{bmatrix} 0 & 0 & 0 \\ 0 & 0 & -a=0.3777 \\ 0 & -b=-0.1984 & 0 \end{bmatrix}$ | $\begin{bmatrix} 0 & 0 & a=-0.3777 \\ 0 & 0 & 0 \\ b=0.1984 & 0 & 0 \end{bmatrix}$ | $\begin{bmatrix} 0 & c=0.1277 & 0 \\ -c=-0.1277 & 0 & 0 \\ 0 & 0 & 0 \end{bmatrix}$ |
| RuO$_2$ (101) *x*[010] *y*[$\bar{1}$01] | $\mathcal{T}$-odd $\Gamma \approx 25$ meV | $\begin{bmatrix} 0 & 0 & 0 \\ 0 & -0.3650 & -0.5756 \\ 0 & 0.2977 & 0.3650 \end{bmatrix}$ | $\begin{bmatrix} 0 & -2.1367 & 0.6457 \\ -1.7286 & 0 & 0 \\ 1.2377 & 0 & 0 \end{bmatrix}$ | $\begin{bmatrix} 0 & -2.2722 & 2.1367 \\ -2.5536 & 0 & 0 \\ 1.7286 & 0 & 0 \end{bmatrix}$ |
| | $\mathcal{T}$-odd $\Gamma \approx 50$ meV | $\begin{bmatrix} 0 & 0 & 0 \\ 0 & -0.1976 & -0.2930 \\ 0 & -0.1472 & 0.1916 \end{bmatrix}$ | $\begin{bmatrix} 0 & -1.0940 & 0.3290 \\ -0.8882 & 0 & 0 \\ 0.6274 & 0 & 0 \end{bmatrix}$ | $\begin{bmatrix} 0 & -1.1616 & 1.0940 \\ -1.3034 & 0 & 0 \\ 0.8882 & 0 & 0 \end{bmatrix}$ |
| | $\mathcal{T}$-even | $\begin{bmatrix} 0 & 0 & 0 \\ 0 & 0.0838 & 0.3200 \\ 0 & -0.2562 & -0.0838 \end{bmatrix}$ | $\begin{bmatrix} 0 & -0.1168 & -0.2972 \\ 0.0330 & 0 & 0 \\ 0.1756 & 0 & 0 \end{bmatrix}$ | $\begin{bmatrix} 0 & 0.2083 & 0.1168 \\ -0.1505 & 0 & 0 \\ -0.0330 & 0 & 0 \end{bmatrix}$ |

**Table S1**: Calculated SHC for RuO$_2$ films in units of $10^5 \left(\frac{\hbar}{2e}\right)(\Omega\, m)^{-1}$. The components with grey background only emerge in the presence of SOI.

From symmetry considerations the most general dependence of the SHC on the angle $\psi$ for a (101)-oriented sample should be:

| SHC | $\mathcal{T}$-odd | $\mathcal{T}$-even |
|---|---|---|
| $\sigma_{zx}^x$ | $\sin\psi \cos\psi \, [-2B\cos^2\theta_N + (A+C)\sin^2\theta_N]$ | $(a+c)\sin^2\theta \sin\psi \cos\psi$ |
| $\sigma_{zx}^y$ | $\sin^2\psi \, (B\cos^2\theta_N - A\sin^2\theta_N)$ $+ \cos^2\psi \, (-B\cos^2\theta_N + C\sin^2\theta_N)$ | $\sin^2\psi \, (b\cos^2\theta_N - a\sin^2\theta_N)$ $+ \cos^2\psi \, (b\cos^2\theta_N + c\sin^2\theta_N)$ |
| $\sigma_{zx}^z$ | $(B+C)\sin\theta_N \cos\theta_N \cos\psi$ | $(-b+c)\sin\theta \cos\theta \cos\psi$ |

Here A, B, C are the non-zero elements of the $\mathcal{T}$-odd SHC tensors and a, b, c are the non-zero elements of the $\mathcal{T}$-even SHC tensors of (001)-oriented RuO$_2$ as discussed in Table S1. $\theta_N$ is the tilt angle of the Néel vector relative to the [101] (out of plane) axis. $\psi$ is the angle between $J_C$ and [010] axis in the (101) plane.

Since $B \ll A, C$ we can accurately approximate the $\psi$ dependence the $\mathcal{T}$-odd SHC as:

$$\sigma_{zx}^y \propto (A+C)\sin^2\theta_N \cos^2\psi - A\sin^2\theta_N \tag{13}$$

$$\sigma_{zx}^z \propto C\cos\theta_N \sin\theta_N \cos\psi \tag{14}$$

$$\sigma_{zx}^x \propto (A+C)\sin^2\theta_N \cos\psi \sin\psi, \tag{15}$$

In the absence of SOI, *A* is vanishing and hence these expressions would correspond to generation of spin current with a spin orientation exactly parallel to the Néel vector, $\mathcal{N}$. However, since Ru is a heavy metal with significant SOI, we predict A/C ~ 0.2. The predicted spin Hall conductivities then correspond to a spin angle $\theta_S$ with $\tan\theta_S = [(A+C)/C]\tan\theta_N$ so that for (101) RuO$_2$ $\theta_S$ is misaligned from $\theta_N$ by about 5°. This is consistent with what we measure experimentally, as described in the main text.

The predicted $\mathcal{T}$-odd SHC increases by about a factor of 2 as the broadening parameter Γ is decreased from 50 to 25 meV. This suggests that the $\mathcal{T}$-odd SHC should also be sensitive to temperature, increasing as the temperature is lowered due to an increasing electron scattering time.

For a single-domain RuO$_2$ antiferromagnet, we predict that the $\mathcal{T}$-odd SHC is more than an order of magnitude larger than the $\mathcal{T}$-even SHC. However, stable domains can have the Néel vector pointing along either [001] or [00$\bar{1}$], and all components of $\mathcal{T}$-odd SHC will flip sign if $\mathcal{N}$ is reversed. If the sample contains an exactly balanced distribution of [001] and [00$\bar{1}$] the $\mathcal{T}$-odd SHC should go to zero. However in the presence of any imbalance in the domain distribution, both $\mathcal{T}$-odd and $\mathcal{T}$-even SHCs can be non-zero, with qualitatively-similar dependences on the angle of applied current $\psi$.

## S5. Exchange bias in (101)RuO$_2$/Py samples

In magnetization hysteresis measurements as a function of in-plane applied magnetic field done at room temperature on (101)RuO$_2$(6 nm)/Py(4 nm) films, when the magnetic field is swept parallel to the projection of the Néel vector in the sample plane (the [$\bar{1}$01] axis) we observe abrupt switching with a small coercive field and saturation field ≈ 2.5 mT (Fig. S5(c)), whereas field sweeps in the perpendicular direction [010] give gradual, magnetic-hard-axis behavior with a saturation field ≈ 20 mT (Fig. S5(d)). This can be compared hysteresis loops for (101)IrO$_2$(6 nm)/Py(4 nm) samples in which the IrO$_2$ is isostructural to RuO$_2$ but lacks antiferromagnetism (Fig. S5(e,f)). For the IrO$_2$/Py samples, both directions of magnetic field give abrupt switching with small coercive fields. All measurements of spin-transfer torque in the main text are conducted using applied fields much larger than 20 mT.

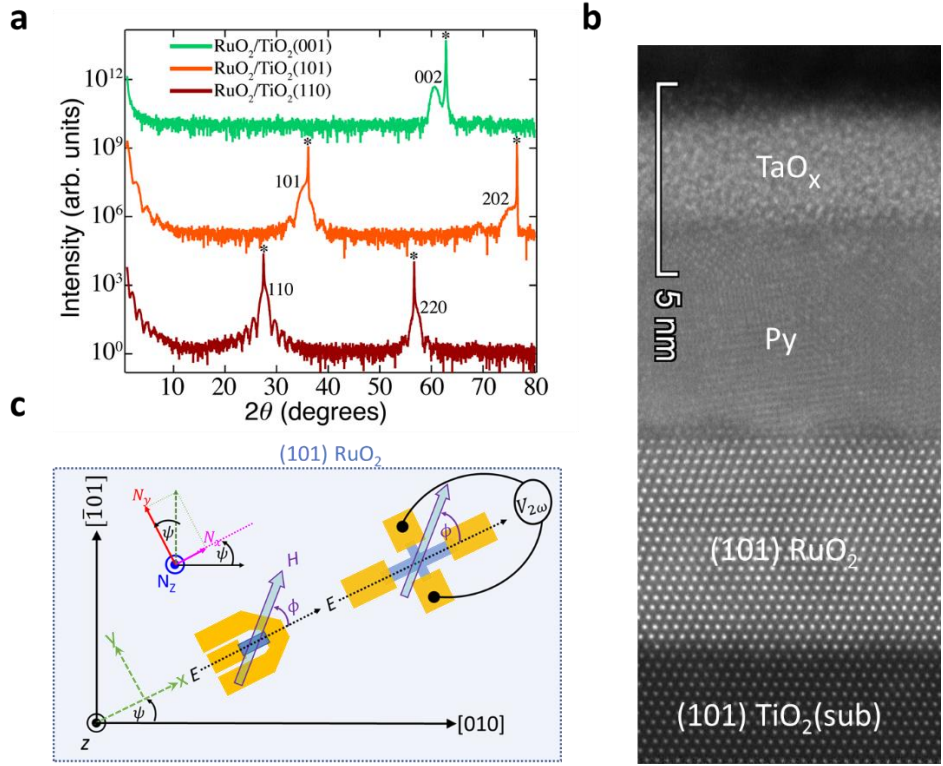

Fig. S1. (a) θ-2θ X-ray diffraction measurements for $RuO_2$ films grown on (110) $TiO_2$, (101) $TiO_2$, and (001) $TiO_2$ substrates. The asterisks (*) indicate the location of the substrate peaks, and the indices of the $RuO_2$ film peaks are included in the plot for each sample. (b) High resolution annular dark-field scanning transmission electron microscope image of a (101) $RuO_2$/Py bilayer. (c) Schematics of the experimental devices (both ST-FMR and HH devices) as fabricated on (101)-oriented $RuO_2$ films.

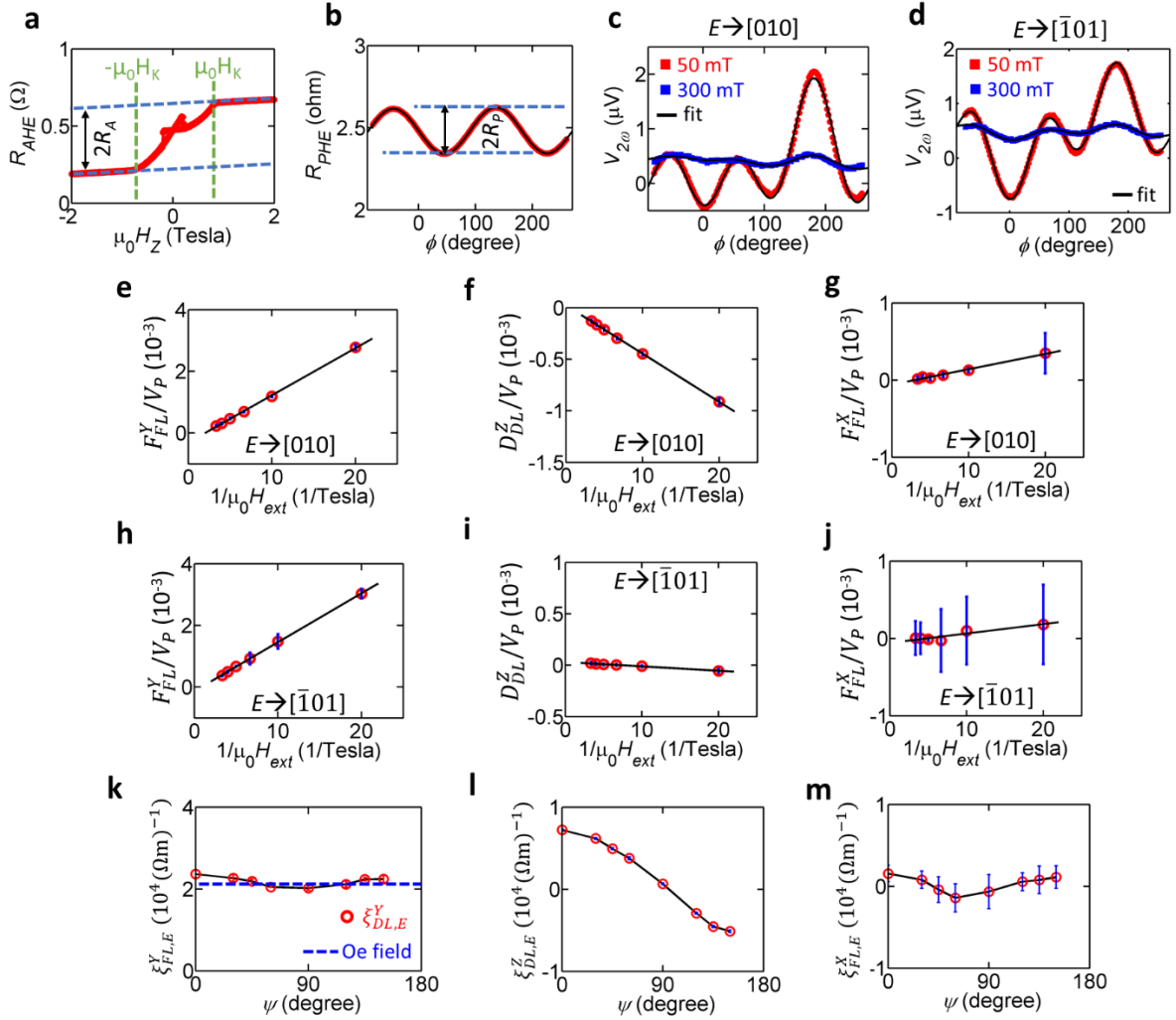

Fig. S2. Measurements for (101)RuO$_2$(6 nm)/Py(4 nm) Hall bars. (a) Hall resistance as a function of out-of-plane magnetic field. (b) Hall resistance as a function of in-plane magnetic-field angle $\phi$ for a field magnitude $\mu_0 H$ = 100 mT. (c) Second-harmonic Hall voltage for an applied electric field along the [010] direction ($\psi$=0°) for two different magnitudes of magnetic field. (d) Second-harmonic Hall voltage for an applied electric field along the [$\bar{1}$01] direction ($\psi$=90°) for two different magnitudes of magnetic field. (e,f,g) Normalized second-harmonic Hall voltages for $\psi$=0° as a function of the inverse of the applied magnetic field, used to calculate $\xi^Y_{FL,E}$, $\xi^Z_{DL,E}$ and $\xi^X_{FL,E}$ for $\psi$=0. (i,j,k) Normalized second-harmonic Hall voltages for $\psi$=90° as a function of the inverse of the applied magnetic field, used to calculate $\xi^Y_{FL,E}$, $\xi^Z_{DL,E}$ and $\xi^X_{FL,E}$ for $\psi$=90°. (k) Symbols: measured values of $\xi^Y_{FL,E}$ from the second-harmonic Hall measurements Dashed line: estimated contribution from the Oersted field. (l,m) $\xi^Z_{DL,E}$ and $\xi^X_{FL,E}$ determined from the second-harmonic Hall measurements for different values of $\psi$. Solid lines in (k,l,m) are guides to the eye.

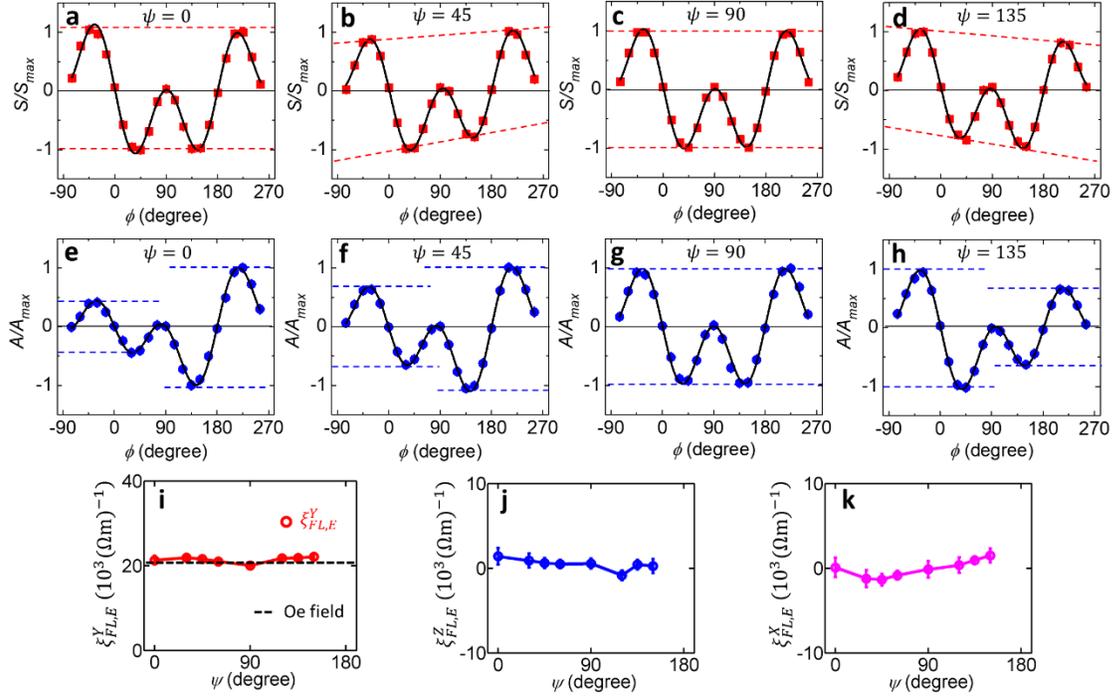

Fig. S3. Results of ST-FMR measurements for (101)RuO$_2$(6 nm)/Py(4 nm) samples. (a-d) Angular dependence of normalized symmetric component $S/S_{max}$ for electric fields applied at different angles ($\psi$) with respect to the [010] direction. (e-h) Angular dependence of normalized antisymmetric component $A/A_{max}$ for electric fields applied at different angles ($\psi$) with respect to the [010] direction. Red and blue dashed lines are to guides the eye to understand the differences in the figures. (i-k) Estimated value of the field-like torque efficiency per unit field as a function of $\psi$ for effective fields in the $Y$, $Z$ and $X$ directions.

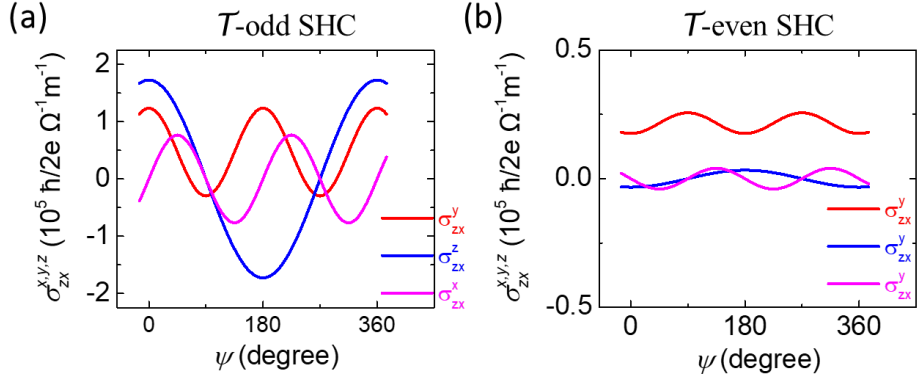

Fig. S4. Predicted dependence on the current angle $\psi$ for the (a) $\mathcal{T}$-odd and (b) $\mathcal{T}$-even spin Hall conductivity in a single-domain (101)-oriented RuO$_2$ film.

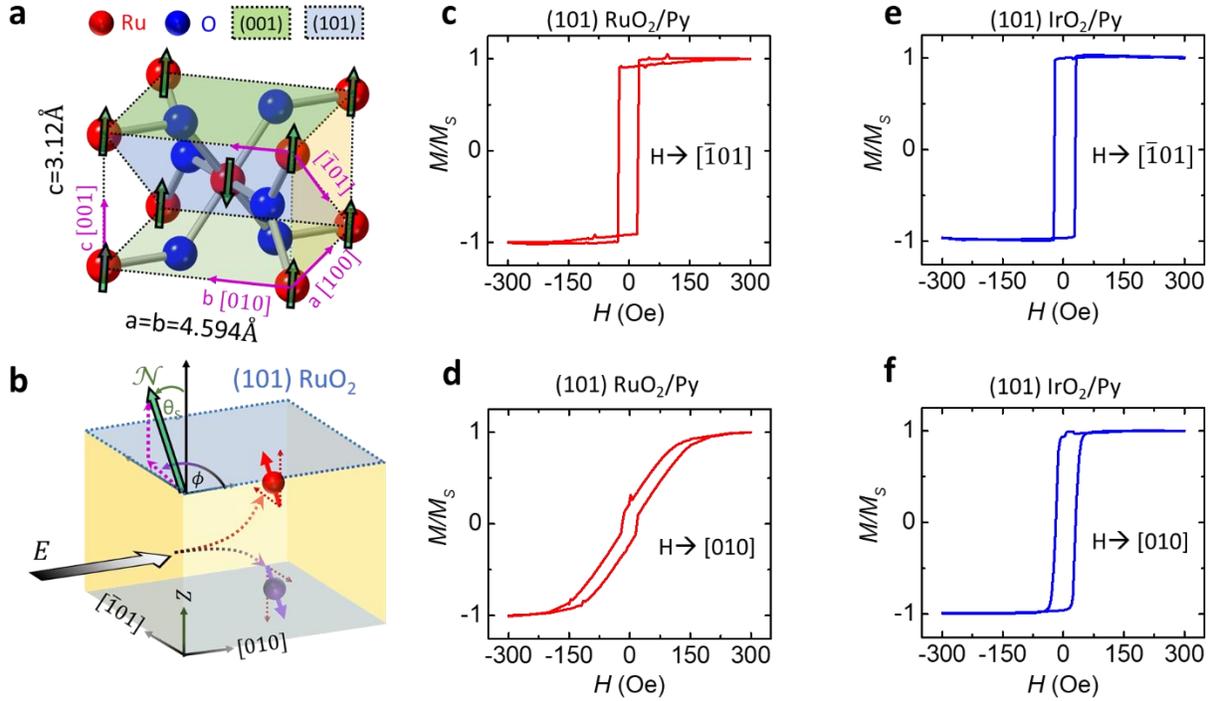

Fig. S5. (a) Schematic of the spin configuration (green arrows) for (001)-oriented RuO$_2$. (b) Schematic showing the orientation of the Néel vector for (101)-oriented RuO$_2$. (c,d) Magnetization versus applied magnetic field hysteresis loops for (101) RuO$_2$ films (6 nm) for magnetic field swept (c) along the [$\bar{1}$01] axis and (d) along the [010] axis. (e,f) Magnetization versus applied magnetic field hysteresis loops for (101) IrO$_2$ films (6 nm) for magnetic field swept (e) along the [$\bar{1}$01] axis and (f) along the [010] axis.